\documentclass[prb,twocolumn,superscriptaddress,showpacs]{revtex4}
\usepackage[dvips]{graphicx}
\usepackage{latexsym}
\newcommand{\fig}[2]{\includegraphics[width=#1]{#2}}

\begin{document}

\renewcommand{\ni}{{\noindent}}
\newcommand{\dprime}{{\prime\prime}}
\newcommand{\be}{\begin{equation}}
\newcommand{\ee}{\end{equation}}
\newcommand{\bea}{\begin{eqnarray}} 
\newcommand{\eea}{\end{eqnarray}}
\newcommand{\nn}{\nonumber} 
\newcommand{\bk}{{\bf k}}
\newcommand{\bQ}{{\bf Q}}
\newcommand{\bN}{{\bf \nabla}}
\newcommand{\bA}{{\bf A}}
\newcommand{\bE}{{\bf E}}
\newcommand{\bj}{{\bf j}}
\newcommand{\bJ}{{\bf J}}
\newcommand{\bs}{{\bf v}_s}
\newcommand{\bn}{{\bf v}_n}
\newcommand{\bv}{{\bf v}} 
\newcommand{\la}{\langle}
\newcommand{\ra}{\rangle} 
\newcommand{\dg}{\dagger}
\newcommand{\br}{{\bf{r}}} 
\newcommand{\brp}{{\bf{r}^\prime}} 
\newcommand{\bq}{{\bf{q}}}
\newcommand{\hx}{\hat{\bf x}} 
\newcommand{\hy}{\hat{\bf y}}
\newcommand{\bS}{{\bf S}} 
\newcommand{\cU}{{\cal U}}
\newcommand{\cD}{{\cal D}} 
\newcommand{\bR}{{\bf R}}
\newcommand{\pll}{\parallel} 
\newcommand{\sumr}{\sum_{\vr}} 
\newcommand{\cP}{{\cal P}} 
\newcommand{\cQ}{{\cal Q}} 
\newcommand{\cS}{{\cal S}}
\newcommand{\upa}{\uparrow} 
\newcommand{\dna}{\downarrow}

\title{Testing for topological order 
in variational wavefunctions for $Z_2$ spin liquids}
\author{Arun Paramekanti}
\affiliation{Department of Physics and
Kavli Institute for Theoretical Physics, University 
of California, Santa Barbara, CA 93106--4030}
\author{Mohit Randeria}
\affiliation{Department of Theoretical Physics, Tata Institute of 
Fundamental Research, Mumbai 400 005, India}
\affiliation{Department of Physics, University of Illinois at 
Urbana-Champaign, IL 61801}
\author{Nandini Trivedi}
\affiliation{Department of Theoretical Physics, Tata Institute of 
Fundamental Research, Mumbai 400 005, India}
\affiliation{Department of Physics, University of Illinois at 
Urbana-Champaign, IL 61801}
\begin{abstract}
\vspace{0.1cm}
We determine the conditions under which a spin-liquid Mott insulator
$\vert 0\ra$ defined by a Gutzwiller projected BCS state at half-filling
is $Z_2$ fractionalized. We construct a trial vison
($Z_2$ vortex) state $\vert V\ra$ by projecting an $hc/2e$ vortex threading 
the hole of a cylinder/torus and examine 
its overlap with $\vert 0\ra$ using analytical and numerical calculations.
We find that generically the overlap vanishes in the thermodynamic limit,
so the spin-liquid is $Z_2$ fractionalized. We point out the relevance
of these results to numerical studies of Hubbard-like models and
spin models which have been 
recently reported to possess spin liquid phases. We also
consider possible implications for flux-trapping experiments that have
tested for $Z_2$ fractionalization in underdoped high 
temperature superconductors.
\typeout{polish abstract}
\end{abstract}

\maketitle
\section{introduction} 

Spin liquid Mott insulators have long been viewed as candidates for 
the ground state of certain frustrated magnetic systems \cite{fazekas}
in dimensions
$D \geq 2$. They have been of interest in the context of high 
temperature superconductivity following Anderson's proposal 
\cite{anderson} that doping 
such insulators may be a novel route to superconductivity.
Progress in our theoretical understanding of a certain  class
of spin liquids has come from two lines of attack over the past
decade. First, 
working from the point of effective theories, it has been shown
that spin-liquid insulators in dimensions $D \geq 2$ may emerge as 
deconfined
quantum phases of $Z_2$ gauge theories \cite{sachdev,wen,senthil_1}
at zero temperature.
In this phase, an $S=1$ excitation, which is the elementary excitation
of a conventional
magnet, may break up into two $S=1/2$ particles, called {\it spinons},
which are minimally coupled to an Ising gauge field; hence also the term 
``$Z_2$ fractionalized spin liquids'' for such insulators. 
The more recent line of progress has been in the construction of 
simple model Hamiltonians which have no special gauge symmetries
but which can nevertheless be shown to possess such
$Z_2$ fractionalized phases --- these include quantum dimer models on
the triangular \cite{moessner} and Kagome \cite{misguich} lattices, and 
certain Hubbard-like boson models \cite{balents,motrunich}. 

There are a few experimental indications of spin liquid states 
in two dimensions. Recent low temperature NMR and 
susceptibility measurements \cite{kanoda} in a quasi-2D organic insulator
indicate no magnetic phase transitions down to $40 mK$.
Similarly, no ordering transitions are found in Helium-3
adsorbed on graphite\cite{ishimoto} even at very low temperatures 
around $10 \mu K$. These temperatures are nearly two orders of magnitude
smaller than the estimated magnetic exchange energy scale in
these respective systems.
Further, experimental signatures of spinons have been reported in neutron 
scattering experiments\cite{coldea} on Cs$_2$CuCl$_4$.

In the past few years, there has also been a lot of numerical work
on microscopic models, which are not analytically tractable, searching
for spin-liquid phases.
In particular, exact diagonalization studies of a multiple-spin exchange 
model on a triangular
lattice \cite{lhuillier}, and Monte Carlo studies of
certain two-dimensional Hubbard-like models
\cite{imada,assaad} indicate that such models
may possess, in a regime of parameters, insulating phases with 
no obvious broken translational or broken spin rotational 
symmetries.

What kind of spin liquids could these
experiments and numerical studies be probing? 
In order to show how to
answer this question in the context of numerical
studies, we focus on ``topological order'' which is a sharp test
of whether the phases being accessed in the
numerics are $Z_2$ fractionalized spin liquids.
Topological order refers to degeneracies in the spectrum of the 
Hamiltonian which depend on the topology of the manifold on which
the system lives.
This may be understood as follows:
if the system is in the deconfined phase
of a $Z_2$ gauge theory, it necessarily implies a novel gapped excitation 
in the bulk of the system
\cite{read,wen,senthil_1}, namely the Ising vortex of
the $Z_2$ gauge field called the ``vison''.
However, a vison threading noncontractible loops 
of the manifold, i.e. threading the holes of the cylinder or torus, 
does not cost any energy in the thermodynamic limit. This leads 
to a topological degeneracy of the low energy eigenstates 
\cite{read,wen,senthil_1}, which may be viewed as states
with/without visons threading the holes of the cylinder or torus.
In summary, for a $Z_2$ fractionalized spin-liquid insulator
we expect a two-fold degeneracy of the low energy spectrum 
on a cylinder and a four-fold degeneracy on a torus.
If indeed the above
numerical studies are probing $Z_2$ fractionalized spin-liquid 
phases, it would be useful to check if they possess topological
order consistent with this phase.\cite{footnotelutt}

In this paper, we address the following issues:
(a) How can one think about the vison in terms of {\it electronic}
coordinates?
(b) Under what conditions is the vison well-defined,
leading to $Z_2$-fractionalization and topological degeneracy 
in spin-liquid insulators? We show
how one may obtain an 
estimate of the length scale beyond which fractionalization 
is apparent, consider its implications for the cuprate
superconductors, and also point out how
one can use the method described here to test for $Z_2$ topological 
order in other numerical studies of Hubbard-like models, 
superconductors. Our work builds upon and extends
some of the results of Ivanov and Senthil \cite{ivanov}.

Throughout this paper, we will discuss the above issues in the
context of a specific spin-liquid state, namely a Gutzwiller
projected d-wave BCS wavefunction on a square lattice.
This wavefunction has been shown to provide a remarkably good
description of the superconducting state of the high Tc cuprates
over a wide range of doping \cite{paramekanti}, and the present study 
was carried out
to examine its implications for the undoped insulator and explore
connections with
phenomenological theories of the cuprates involving $Z_2$ 
fractionalization. However, our study serves more generally to 
illustrate the method of
identifying $Z_2$ spin-liquid states either using Gutzwiller
projected variational wavefunctions, or even other 
numerical methods as we discuss in Section \ref{othermethods}.

\section{Constructing the vison wavefunction}

We begin with a discussion of how to construct a vison wavefunction
in terms of microscopic electronic degrees of freedom, which form the
basis of the numerical studies.
For conceptual clarity, consider first a square lattice of
$L^2$ sites ``wrapped'' into a cylinder along the  $\hat{x}$ direction.
The spin-liquid ground states $\vert 0 \ra$ of interest to us are
given by the Gutzwiller projection
of $N$-particle d-wave BCS states {\em at half-filling} ($N = L^2$):
\be
\vert 0 \ra \equiv \cP \vert BCS \ra = \cP 
\big[ \sum_{\br,\brp} \varphi^{\vphantom\dagger}(\br - \brp)
c^{\dagger}_{\br\upa} c^{\dagger}_{\brp\dna} \big]^{N/2}
\vert {\rm vac} \ra.
\label{gs}
\ee
Here the pair wavefunction is given explicitly by
\be
\varphi(\rho)= L^{-2} \sum_\bk \exp(i\bk\cdot\rho) 
[\Delta_\bk/(\xi_\bk+\sqrt{\xi_\bk^2 + \Delta_\bk^2})]
\label{varphi}
\ee
with $\bk$'s are chosen consistent with periodic boundary condition
(PBC) along $\hat{x}$. 
The pair wavefunction is parametrized in terms of two
variational parameters \cite{notation} $\mu$ and $\Delta$
which determine $\xi_\bk = \epsilon(\bk) - \mu$,
with $\epsilon(\bk) = -2 t\left(\cos k_x + \cos k_y\right) - 
4 t^\prime \cos k_x \cos k_y$ 
and $\Delta_\bk = \Delta \left(\cos k_x - \cos k_y\right)/2$.
The Gutzwiller projection operator 
$\cP=\prod_{\br} (1-n_{\br\upa} n_{\br\dna})$ restricts the
configurations to have exactly one electron per site.

While the state before projection describes a
superconductor, the projected state is an insulator with one electron
(or equivalently one spin) per site, and no long range magnetic order
\cite{paramekanti}.

To test for $Z_2$ fractionalization, we would like to construct a 
trial state for a vison threading a hole of the cylinder to check
for topological order.
We do this in two steps. First, we 
construct a wavefunction for an $hc/2e$ vortex threading
the hole of the cylinder. Next, we 
we Gutzwiller project this $hc/2e$ vortex state and
present four arguments for why gives us a good 
candidate for a vison state.

To construct a superconducting state with one $hc/2e$ vortex threading
the hole of the cylinder, we modify the pairing 
function in the BCS ground state as
\be
\varphi^{\vphantom\dagger}(\br - \brp) \to
\exp[i\bQ\cdot(\br + \brp)/2]\varphi_{A}(\br - \brp),
\label{vortexstate}
\ee
where 
$\bQ = 2\pi\hat{x}/L$. This gives a center of mass momentum
$\bQ$ to each Cooper pair and twists the phase of the condensate 
by $2\pi$ going around the cylinder.
However, the requirement of a single-valued wavefunction
when even {\it one} of the electrons of the Cooper pair 
($\br$ or $\brp$) goes around the cylinder, constrains
$\bk$'s to satisfy {\em anti}periodic boundary 
conditions (APBC) along $\hat{x}$ in the Fourier transform
in defining $\varphi_{A}(\br-\brp)$ via equations analogous to 
Eqs.(\ref{gs},\ref{varphi}); hence the subscript $A$ on
$\varphi_{A}(\br - \brp)$ in Eq.~\ref{vortexstate} above.
Clearly the state with one $hc/2e$ vortex has a current flow
and it can also be written as
\bea
\vert hc/2e\ra_y =
\exp\big(i\sum_\br \frac{\bQ\cdot\br}{2}[n_\upa(\br) + n_\dna(\br)]\big) \nn\\
\times \big[ \sum_{\br,\brp} \varphi_{A}^{\vphantom\dagger}(\br - \brp)
c^{\dagger}_{\br\upa} c^{\dagger}_{\brp\dna} \big]^{N/2}
\vert {\rm vac} \ra.
\label{hcby2e}
\eea

Gutzwiller projecting this state at half-filling
fixes $n_\upa(\br)+n_\dna(\br)=1$, and the prefactor multiplying
the wavefunction in Eq.(\ref{hcby2e}) then drops out as a trivial phase
factor since it is
independent of the spin configuration. This gives
\be
\vert V_y \ra \equiv \cP\vert hc/2e\ra_y = \cP
\big[ \sum_{\br,\brp} \varphi_{A}^{\vphantom\dagger}(\br - \brp)
c^{\dagger}_{\br\upa} c^{\dagger}_{\brp\dna} \big]^{N/2}
\label{vison}
\ee
Thus the Gutzwiller projected $hc/2e$ vortex state
$\vert V_y \ra$ simplifies to a form identical to Eq.~(\ref{gs}) 
with $\varphi^{\vphantom\dagger} \rightarrow \varphi_{A}$, and
corresponds to imposing antiperiodic boundary conditions on the
electrons before projecting them. 

Why is this a good candidate
for a vison state?
First, a projected $hc/e$ vortex with $\bQ = 4\pi\hat{x}/L$ 
does not require any change in BC's on $\varphi$
to ensure single-valuedness, and the phase factors again drop out
at half-filling. It is thus identical to the ground state 
$\vert 0 \ra$.
Hence we only have two kinds of vortex states upon projection: all 
superconducting states with an 
even number of $hc/2e$ vortices threading the cylinder collapse
onto the ground state $\vert 0\ra$, those with an odd number
of $hc/2e$ vortices threading the cylinder collapse onto 
$\vert V_y\ra$. This renders manifest the $Z_2$ character of
a projected $hc/2e$ vortex --- two vortices is the same as none
--- which makes them attractive candidates for visons.

Second, we can look at the crystal momentum of the projected vortex. 
The vortex state before projection has a crystal momentum $\bQ$ per 
singlet pair, or a total crystal momentum of $\bQ N/2$. Since
the Gutzwiller projection operator commutes with the translation
operator, the crystal momentum is unchanged by projection. Thus for
a system with dimensions $L_x \times L_y$, the projected vortex has a 
total crystal momentum $P_x = \pi L_y$. Since the crystal momentum
is only defined modulo $2\pi$, this means the projected $hc/2e$ vortex
wavefunction carries momentum $\pi$ for odd-$L_y$ and {\it no momentum}
for even-$L_y$. As shown 
elsewhere \cite{ashvin}, this is precisely the momentum of a vison in
an insulator
described by a $Z_2$ gauge theory with one spin per site. Thus the
momentum quantum number of the projected $hc/2e$ vortex is consistent
with that of the vison. As discussed below, we will always work with
even-$L_y$ in this paper.

Third, one can consider a specific limit of the pair wavefunction,
namely $\mu \to -\infty$, which corresponds to having singlet pairs
only on neighboring sites, like in a nearest neighbor dimer model.
The state $\vert 0\ra$ corresponds to a superposition of nearest
neighbor dimer (singlet) configurations.
In this limit, due to the antiperiodic boundary conditions on
$\varphi_{A}(\br -\br')$,
it is straightforward to show that every 
configuration in which an odd number of singlets lie on links such 
that one spin of the singlet is at a site $(L_x,y)$
and the other spin lies at $(1,y)$ (taking into account all allowed 
$y$) has an amplitude which differs in sign from its amplitude
in the state $\vert 0\ra$.
This is however identical to the well-known 
construction of the $Z_2$ vison threading a cylinder in the context 
of dimer models \cite{kivelson,read}.

Finally, in terms of a dual theory of vortices,
the way to obtain a $Z_2$ fractionalized 
insulator starting from a superconducting state is by condensing pairs
of $hc/2e$ vortices \cite{nodalliquid}, 
thus quantum disordering the superconductor. In
this case, the vison is the remnant of the $hc/2e$ vortex which has
not condensed. Since Gutzwiller
projecting a superconductor at half-filling fixes the number of 
particles, it amounts to phase disordering the superconductor, and
this also leads one to suspect that the Gutzwiller projected $hc/2e$ 
vortex is the vison.

Generalizing the above discussion to a torus, $\vert 0 \ra$ corresponds 
to PBC along both $\hat{x}$ and $\hat{y}$, $\vert V_x \ra$ and 
$\vert V_y \ra$ correspond to PBC/APBC and $\vert V_{xy} \ra$ to APBC 
along both directions.
These proposed vison wavefunctions are designed to
investigate the existence of disconnected topological sectors 
as evidence for fractionalization as we discuss in the following
section. A vison excitation in the bulk, on the
other hand, would need to have a $Z_2$ flux piercing the plane
of the lattice and would require careful consideration of the
superconducting vortex core prior to projection, which we avoid 
here. 


\section{How do we test for topological order?}

A SC state with a vortex threading the hole of
the cylinder/torus carries current and is orthogonal to the ground state.
However, Gutzwiller projecting the BCS wavefunction at 
half-filling destroys SC order \cite{paramekanti} and results in an 
translationally invariant insulator \cite{footnote2}. 
The key question then is:
{\em does any remnance of vorticity survive projection?}
If it does, then
$\la V_\alpha \vert 0 \ra\!\! =\!\! \la V_\alpha \vert V_\beta \ra
\!\!= \!\!0$,
for $\alpha,\!\!\beta\!\! =\!\! x,y,xy$, the vison is well-defined
leading to topological degeneracy
\cite{footnote3}, 
and one is in a $Z_2$ fractionalized
phase. If, on the other hand, the proposed vison state has nonzero 
overlap with the state $\vert 0\ra$ in the thermodynamic limit,
one is in a conventional unfractionalized phase.

Since we have shown above that the vison state on a cylinder
with odd-$L_y$ carries crystal momentum $\pi$,
it is trivially orthogonal to the ground state $\vert 0\ra$ even on
a finite system. However
this by itself does not tell us anything about topological order since
if this state becomes degenerate with the ground state it may mix 
with the ground state in the thermodynamic limit and break translational
symmetry --- for instance, this is what is known to happen in
{\it any} gapped phase of
an odd number of coupled spin chains (an ``odd-leg ladder''), 
as is well-known from an extension\cite{Affleck}
of the Lieb-Schultz-Mattis theorem in one dimension.\cite{LSM} 
However, in the limit of infinite number of coupled chains, the 
order parameter in the broken symmetry state may survive
leading to a conventional broken symmetry state in two dimensions,
or it may vanish leading to a translationally invariant 
two-dimensional spin liquid \cite{capriotti}.
These two possibilities cannot be distinguished by an overlap 
calculation with odd-$L_y$.

For purposes of testing for topological order, 
we are therefore really 
interested in the case of even-$L_y$ where the trial vison
state carries no crystal momentum, and it is thus not trivially orthogonal to 
$\vert 0\ra$ on finite-size systems. Showing that the overlap between
the ground state and the proposed vison state vanishes
in the thermodynamic limit on cylinders or torii with
even length in each direction is therefore a crucial step in establishing
topological order.

\begin{figure}
\begin{center}
\vskip-2mm
\hspace*{0mm}
\centerline{\fig{3.0in}{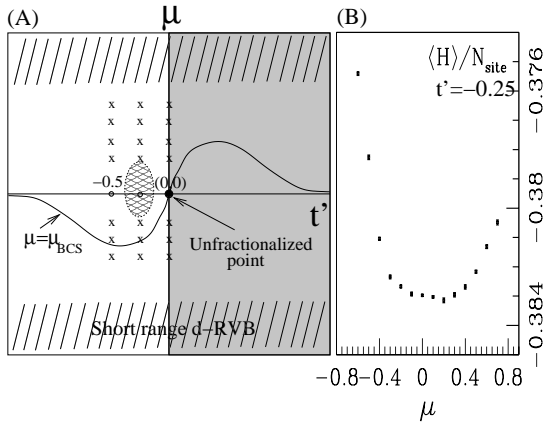}}
\vskip-6mm
\caption{(A) ``Phase diagram'' showing that $\vert 0\ra$ is $Z_2$ 
fractionalized in the entire $(t^\prime,\mu)$-plane except
at the bipartite symmetric point $(0,0)$ (see text for details). (1) 
Shaded and unshaded half-planes
are related by symmetry $(t^\prime,\mu) \to (-t^\prime,-\mu)$;
(2) hatched region for large negative $\mu$ is the 
short-range RVB (resonating valence bond)
limit; (3) analytical arguments for
$\mu=\mu_{\rm BCS}$ (schematic line) are described in the text;
(4) crosses indicate some points where we
numerically compute overlaps; (5) shaded ellipse around $(-0.25,0)$ 
is the region relevant for high Tc SCs. 
(B) Energy of $\vert 0 \ra$ for the
Hubbard model at half-filling for parameters relevant to cuprates 
($U/t\!\! =\!\! 12, t^\prime/t\!\! =\!\! -0.25$ and optimal $\Delta\!\! 
=\!\! 1.25t$) versus $\mu/t$. The optimal value is
$- 0.3 \lesssim \mu/t \lesssim 0.3$ within statistical errors.}
\label{phasediag}
\end{center}
\vskip-8mm
\end{figure}

\section{``Phase diagram'' and symmetries}

To determine the conditions under which the vison survives, 
it is useful to consider the full parameter space for 
projected BCS states at half-filling, which is the  
$(t^\prime,\mu)$ plane with $t = 1$ and 
$\Delta$ held fixed at some nonzero value
in order to describe a RVB liquid of singlet pairs.
Studying this general class of states will be useful since one 
can check if there are
states which are not $Z_2$ fractionalized in some regime of
parameters,
and ask how the transition from a fractionalized regime to an
unfractionalized regime is reflected in the overlap of the
wavefunctions $\la V_\alpha\vert V_\beta\ra$.

We use symmetry arguments to show that not all
states in this space are distinct upon projection and
$\vert 0 (t^\prime,\mu)\ra = \vert 0 (-t^\prime,-\mu)\ra$. 
In brief, we can change $t^\prime\to -t^\prime$, $\mu \to 
-\mu$ by a global particle-hole transformation 
$c^{\vphantom\dagger}_{\br\sigma} \to (-1)^{x+y} c^{\dagger}_{\br\sigma}$ 
in the wavefunction, also redefining the $\vert {\rm vac} \ra$ since 
empty sites transform into doubly occupied ones. However, with 
{\it exactly} one particle per site such a particle-hole
transformation
interchanges $\upa$-spins and $\dna$-spins at every site for any 
configuration. Since the BCS wavefunctions are spin-singlets this 
leaves the state invariant. We thus restrict attention to 
$t^\prime \le 0$ in Fig.~\ref{phasediag}(A). We next turn to various limits
in which some analytical progress is possible.

\section{Short-range RVB limit}

Note that large negative $\mu$ with 
$\mu < - 4 (\vert t\vert +\vert t^\prime\vert)$
is the bosonic limit \cite{randeria} of the BCS wavefunction, with
$\varphi(\br -\brp)$ decaying exponentially in real space. 
In this limit $\vert 0 \ra$ may be viewed as a short-range RVB state.
Earlier studies of closely related dimer models 
\cite{kivelson,read,bonesteel} indicate that, for short-range 
singlet bonds there exist 
four topological sectors on a torus, which may 
be straightforwardly related to sectors with/without
visons in the $\hat{x}$, $\hat{y}$ directions. In particular,
the states with a definite ``parity'' in the dimer model
notation correspond to a superposition of the states with and
without a vison \cite{kivelson,read,bonesteel}.

Further, two dimer {\it configurations} from 
different sectors have exponentially small overlap $\sim \exp(-\alpha L)$.
A state is a superposition of many configurations, and {\em assuming} that 
individually small overlaps do not add up coherently,
the overlap of two states from different sectors would vanish
as $L \to \infty$. 

We have numerically checked (see below) that the overlap
vanishes for large $L$ validating the above assumption, 
and implying fractionalization in the short range RVB limit.
Using the symmetry discussed above, we conclude that
states in the region with (positive)
$\mu > 4(\vert t\vert +\vert t^\prime \vert)$
are also fractionalized, even though $\varphi(\br- \brp)$ is 
{\it not} obviously short-ranged here.

\section{Further analytical insights} 

We next focus on the curve $\mu=\mu_{\rm BCS}(t^\prime)$ 
(see Fig.~\ref{phasediag}(A)), where $\mu_{\rm BCS}$ yields half-filling 
for the {\em unprojected} BCS state in the grand canonical 
representation ($not$ fixed particle number representation). 
This {\em unprojected} grand canonical state, which we shall
denote as $\vert BCS\ra_{GC}$,
viewed as the ground state of a BCS Bogoliubov-deGennes (BdG)
Hamiltonian $H_{BdG}$,
is a coherent superposition of number eigenstates sharply peaked 
at the correct mean density. Working with the grand canonical
wavefunction proves convenient since in our discussion below,
we will use certain results valid for the BdG Hamiltonian in order to
infer properties of its ground state wavefunction.
Note that Gutzwiller projecting the grand canonical wavefunction at 
half-filling also picks out the $N = L^2$ contribution with no 
double occupancy.

Let us define local $SU(2)$ gauge 
transformations
\cite{affleck} ${\cU}$, generated by
$T_j^+= c^{\dagger}_{j\upa} c^{\dagger}_{j\dna}$,
$T_j^-= c_{j\dna} c_{j\upa}$,
$T^z_j=\sum_{\sigma}c^{\dagger}_{j\sigma}c^{\vphantom\dagger}_{j\sigma} -1$, 
which mix empty and doubly occupied
sites as an $SU(2)$ doublet, but act trivially
on the projected subspace with {\it exactly} one particle per site.
Gutzwiller projection is then equivalent to projection onto
the $SU(2)$ singlet subspace, and we may write any state
$\cP \vert \Phi \ra = \int_\cU \cU \vert \Phi \ra$,
where the integral is over all group elements ${\cU_\theta}=
\exp(i\sum_j \vec{T}_j\cdot\vec{\theta}_j)$.
We can thus write
\be
\la V \vert 0 \ra = \int_\cU \la (hc/2e) \vert \cU \vert BCS \ra_{GC},
\label{overlap}
\ee
which reduces the problem to computing overlaps of
{\em unprojected} grand canonical states.

For a nonbipartite lattice (e.g., with $t^\prime \ne 0$),
one cannot gauge away the off-diagonal term in the
d-wave BCS-BdG Hamiltonian by any unitary transformation
$\cU H_{BdG} \cU^{-1}$, and the ground state of this
transformed Hamiltonian, $\cU\vert BCS \ra_{GC}$, 
is thus always a SC ground state\cite{ivanov}
(or vortex vacuum) for arbitrary $\cU$.
Then $\la (hc/2e) \vert \cU \vert BCS \ra_{GC} =0$ for arbitrary 
$\cU$, and thus $\la V \vert 0 \ra$ vanishes.
Thus nonbipartiteness is a sufficient condition for
fractionalization in Gutzwiller projected d-wave states 
\cite{previous} when $\mu=\mu_{\rm BCS}$. 
Note that if one can show this result
directly working with the wavefunction $\vert BCS\ra$
in the fixed number representation (rather than $\vert BCS\ra_{GC}$),
without appeal to the BdG 
Hamiltonian, then this result would be valid for general $\mu$
and not restricted to $\mu=\mu_{\rm BCS}$. However, we have not 
found a simple way to show this. 

Let us now turn to the bipartite system
(in our case $t^\prime = 0$), where it is well known \cite{affleck}
that one can gauge away the off-diagonal part of the 
BdG Hamiltonian (provided it does not break time reversal) and
transform the BCS state into a staggered flux state.
Now $\la (hc/2e) \vert \cU \vert BCS \ra_{GC}$ is nonzero
for some choice $\cU$ and one cannot use eq.~(\ref{overlap})
to argue that $\la V \vert 0 \ra$ vanishes. 
Note, this is not a proof that the overlap remains nonzero
in the thermodynamic limit since one may worry that 
this particular $SU(2)$ rotation, which converts the d-wave state 
to the staggered flux state, has zero measure in 
the integral and thus still preserve orthogonality.
Our numerical results, presented below,
however indicate that the overlap does remain nonzero in the
thermodynamic limit when
$t'=0$, so that the wavefunction at this special 
point does not describe a $Z_2$ fractionalized state.

\section{Overlaps within pure $Z_2$ gauge theories} 
\label{gauge}

\begin{figure}
\begin{center}
\vskip-2mm
\hspace*{0mm}
\centerline{\fig{2.5in}{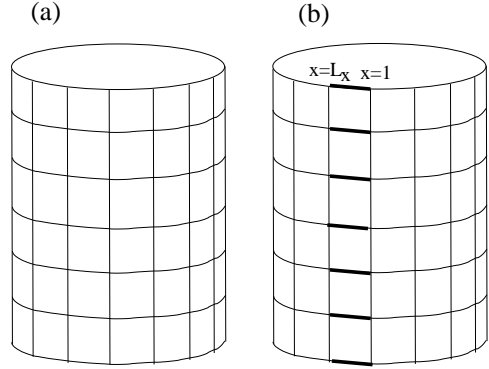}}
\vskip-1mm
\caption{Reference configurations to construct the ground states for
the pure gauge theory with $h=0$ (i.e., no vison dynamics). (A) The 
configuration with $\sigma^z= +1$ on all links. (B) The configuration 
with $\sigma^z=-1$ on the dark bonds, and $\sigma^z=+1$ 
on all other links. These two configurations upon gauge projection 
(see text), lead to the two topologically degenerate ground states 
in the gauge theory with $h=0$.}
\label{cylinder}
\vskip-2mm
\end{center}
\end{figure}

Before proceeding to the numerics, we discuss
how $\la V \vert 0 \ra$ is expected
to scale with system size in an $L_x\times L_y$ system. 
For ease of presentation, we consider the system to be
defined on a cylinder as shown in Fig.~\ref{cylinder}, with
periodic boundary conditions along $\hat{x}$. 
In a unfractionalized phase
the overlap between the states with and without a vison threading
the hole of the cylinder will remain nonzero in the thermodynamic
limit. However, in a fractionalized phase we argue that
on the cylinder the overlap should vanish exponentially with $L_y$.
This is easy to see when the matter fields are gapped 
(e.g., the short-range RVB limit) and may be
integrated out to obtain a $2+1$ dimensional $Z_2$ gauge theory.
Deep in the deconfining phase of the pure gauge theory,
a perturbative calculation yields $\la V \vert 0 \ra \sim 
\exp(-L_y/\xi)$. We obtain this result as follows.

The Hamiltonian for a pure $Z_2$ gauge theory on the square
lattice \cite{kogut} takes the form
\be
H_{Z_2} = - K \sum_{\Box} \prod_{\Box} \sigma^z_{ij} - h \sum_{\la i j\ra} 
\sigma^x_{ij}
\label{hz2}
\ee
where the Pauli matrices $\sigma^x,\sigma^z$ live on the links of the
lattice, and $\Box$ denotes an elementary plaquette on the lattice. 
This Hamiltonian has an extensive number of conserved quantities,
namely the set of operators
\be
\hat{G}(i) = \prod_{j\in i} \sigma^x_{ij},
\ee
where $j\in i$ denotes the set of sites neighboring site-$i$, all commute
with the Hamiltonian, i.e. $[\hat{G}(i),H_{Z_2}] = 0$ for each $i$. Since
$\hat{G}^2(i)=1$, it is clear that $\hat{G}(i)=\pm 1$ are the only
allowed eigenvalues.
For a translationally invariant spin
liquid insulator with exactly one spin per site, it is appropriate to choose
$\hat{G}(i)=-1$ at each site \cite{senthil_1,ashvin}
, and this constrained Hamiltonian is referred
to as an
``odd''-gauge theory \cite{moessner02,ashvin}. 
Let us represent by
$P_{odd}$, the projection into this subspace,
so
\be
P_{odd} \equiv \prod_i \frac{1}{2} [1-\hat{G}(i)].
\ee
Then, any wavefunction $\vert \psi\ra$
may be projected into the ``odd'' gauge theory sector as
$P_{odd} \vert \psi\ra$.

We will now examine the ground states of this Hamiltonian (on a cylinder)
deep in the deconfined phase, which obtains when $h/K \ll 1$. To begin, 
let us consider the extreme limit $h=0$. In this case we want to set the
flux $\prod_\Box \sigma^z = 1$ on each plaquette to minimize the energy
of $H_{Z_2}$ in Eq.~(\ref{hz2}). One ground state in this limit
can be achieved
by setting $\sigma^z=1$ on each link as shown in Fig.~\ref{cylinder}(A),
and acting on this reference
state with $P_{odd}$. Let us call the resulting state $\vert v=0\ra$.
On a cylinder, the system has a second,
distinct, eigenstate $\vert v=1\ra$; this is obtained 
by gauge projecting the reference
state depicted in Fig.~\ref{cylinder}(B), where a column of horizontal links
has $\sigma^z=-1$ such that this reference state also has zero
flux per plaquette. This state $\vert v=1\ra$ can be alternatively 
obtained by acting on the first state $\vert v=0\ra$ with the operator 
\be
V^\dagger = \prod_{i\in column} \sigma^x_{i,i+\hat{x}}
\ee
which may be viewed as the operator which creates a $Z_2$ vortex (vison)
through the hole of the cylinder, since it leads to a change in the 
product of $\sigma^z$ taken around the circumference of the
cylinder from
$+1 \to -1$. We may loosely refer to it as the one-vison state.

Clearly,
\be
\la v=0\vert v=1\ra_{h=0} = 0,
\ee
the overlap between the two states vanishes identically
even for a finite-size system. Since $[V^\dagger,H_{Z_2}]=0$, the
state $\vert v=1\ra$ is an orthogonal eigenstate {\it degenerate}
with the state $\vert v=0\ra$. This leads to the two-fold topological 
degeneracy on a cylinder.

Let us now turn on a small nonzero $h$, so that we are still deep in the
deconfined phase $h/K \ll 1$, and calculate the perturbative
change in each of the two states above. Since $h\sigma^x$ flips the gauge
field on a link from $+1 \to -1$, 
it leads to a superposition within perturbation theory
\be
\frac{\vert +\ra + (h/4K)\vert-\ra}{\sqrt{1+h^2/16 K^2}}
\ee
on each link, and at small $h/K$ the ground state
$\vert v=0\ra_h$ may be built by taking a reference state
which is a direct product 
on all links of the above superposition, and gauge projecting
it using $P_{odd}$. The second ground state $\vert v=1\ra_h$
is obtained by acting with $V^\dagger$ on $\vert v=0\ra_h$. It is 
straightforward
to see that the overlap $\la v=0\vert v=1\ra_h$ vanishes for
odd-$L_y$ whereas it is nonzero for even $L_y$. (This is consistent
with the one-vison state carrying a non-trivial crystal momentum
for odd-$L_y$, which we mentioned earlier.)
We can estimate by a direct calculation that in fact the overlap for
even $L_y$ scales
as $\sim (h/K)^{L_y}$ which 
vanishes exponentially with increasing $L_y$, leading to two orthogonal
states in the thermodynamic limit. Since $[V^\dagger,H_{Z_2}]=0$ even
for nonzero $h$, this second state is an orthogonal {\it ground} state
in the thermodynamic limit. We again recover the topological degeneracy,
and see that the no-vison and one-vison states have an overlap which
vanishes as $\exp(-L_y/\xi)$ for sufficiently large system sizes. 
Similar results have been obtained in Ref.\onlinecite{moessner02}.

A simple way to understand the above results is 
as follows. For a two-state system such as a single-particle in a 
double-well potential with a finite barrier, there is a nonzero 
amplitude within perturbation theory for a particle localized in one 
well to be found in the other well due to tunneling across the
barrier leading to a nonzero overlap between states where the particle
localized in different wells. 
For a collective coordinate such as the vison, a
``string'' of length $L_y$, the tunneling between the two states,
with the vison localized {\it within} the cylinder or {\it outside}
the cylinder, proceeds through 
intermediate states 
where parts of the string lie in the barrier region and the
overlap is thus 
exponentially small in $L_y$. 

The behavior of the overlap with gapless matter fields 
is less well studied though we expect it to further suppress 
tunneling and lead to a smaller overlap; 
numerically (see below) we find clear evidence 
for exponential decay. Note that this decay is very different
from the decay we might expect from the overlap of two
general unrelated many-body states of a system with fixed
density and $L_x L_y$ sites, which would
go as $\sim \exp(-\beta L_x L_y)$ for large system sizes.

\section{Numerical results}
\label{numerics}

\begin{figure}
\begin{center}
\vskip-2mm
\hspace*{0mm}
\centerline{\fig{3.5in}{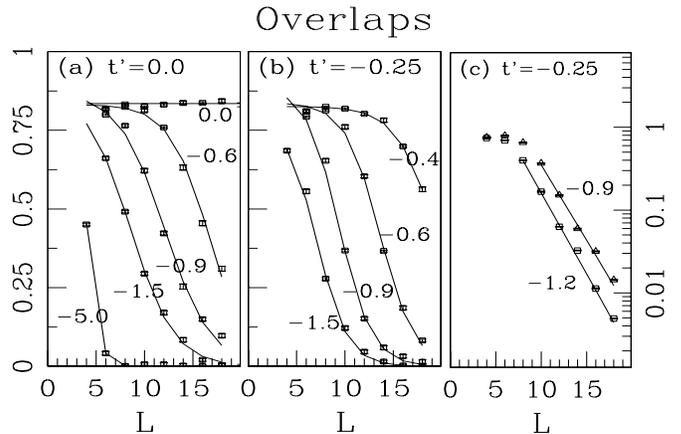}}
\vskip-6mm
\caption{Size dependence of the overlap $\la V_x|V_y\ra (L)$ 
for $\Delta = 1.25$ with
(A) $t^\prime = 0$ and (B) $t^\prime=-1/4$. 
The different curves correspond to overlaps for different 
values of $\mu$ as indicated. An overlap vanishing with
increasing system size indicates
a $Z_2$ fractionalized state, a non-vanishing overlap (e.g.,
the curve in (A) for $\mu=0.0$) indicates a state which is not
$Z_2$ fractionalized. The lines are fits
to the form $a (1-\tanh((L-\xi^\ast)/\xi))$,
which works well for all parameters studied,
and from which we extract $\xi^\ast$. (C) Exponential asymptotic
behavior of the overlap shown on semi-log plot (see text
in Section \ref{gauge} and Section \ref{numerics} for details).
}
\label{fits}
\vskip-8mm
\end{center}
\end{figure}

We finally discuss numerical results for vison overlaps 
using the variational Monte Carlo method.
For computational simplicity \cite{numerics}
we focus on $\la V_x|V_y\ra$.
We choose $t=1$, and fix $\Delta = 1.25$ for all the
calculations.\cite{footnotedelta}
The following results are obtained in regimes where
we have given analytical arguments above.
(i) In the short-range RVB regime 
we find that the wavefunction is $Z_2$ fractionalized; see 
Fig.~\ref{fits}(A) for large negative $\mu$, where the overlap 
vanishes even on fairly small system sizes.
(ii) On the curve $\mu=\mu_{\rm BCS}$ we find
that nonbipartite (i.e., $t'\neq 0$) 
wavefunctions are fractionalized as evinced by overlaps
which vanish with increasing 
system size (e.g., the curve in Fig.\ref{fits}(B) with
$t'=-0.25$ and $\mu=-0.4$). (iii) At
the special point $(t^\prime,\mu) =(0,0)$ with
bipartite symmetry, we indeed find that the overlaps are nonzero,
independent of $L$ (see Fig.~\ref{fits}(A)). Thus the 
vison is not well defined and the system is
{\em not} $Z_2$ fractionalized. Although consistent with
the analytic arguments, the numerical results are important because
the arguments were not rigorous for (i) and (iii).

We next evaluate overlaps at $t^\prime =0.0,-0.25,-0.5$ 
for $-1.5 \lesssim \mu \lesssim 1.5$ which, as we
discuss below, covers the region of interest for possible spin-liquid
insulators in the vicinity of high Tc cuprates.
We show in Fig.~\ref{fits} the overlap $\la V_x \vert V_y \ra$
as a function of system size $L$ for a range of $t^\prime$ and 
$\mu$. We find that at small $L$ the 
overlaps are finite and then cross over on a scale $\xi^\ast$ 
to an asymptotic decay $\exp(- 2 L/\xi)$. This behavior can
be simply described by the functional form 
$\la V_x|V_y\ra (L)=a(1- \tanh((L-\xi^\ast)/\xi))$
which appears to fit the data well \cite{footnote4}.
In particular, this form captures the crossover from a 
constant on small system sizes to an asymptotic 
exponential form apparent in Fig.\ref{fits}(C).

Deep in the fractionalized regime, we can extract 
both $\xi^\ast$ and $\xi$ and we find
$\xi^\ast \simeq 2\xi$. However, it is hard to access the
asymptotic behavior in the region of interest around ($t'=0,\mu=0$), 
though we can still reliably extract the crossover scale $\xi^\ast$.

We plot the inverse length scale $1/\xi^\ast$ for $t^\prime=0,-0.25,-0.5$ at 
different $\mu$ in Fig.~\ref{scaling}. For $t^\prime=0$, we see that
$\xi^\ast \to \infty$ as $\mu \to 0$,
fully consistent with the finite overlap independent of $L$ on 
accessible system sizes for $(t^\prime,\mu) = (0,0)$ seen 
in Fig.~\ref{fits}(A). It also clear that $\xi^\ast$ remains finite 
everywhere away from the origin though it may become quite large in 
its vicinity.

Since fractionalization manifests itself at length scales 
larger than $\xi^\ast$, our numerical results strongly suggest 
that all points in the phase diagram of Fig.~\ref{phasediag}(A) are 
fractionalized, except for the origin where $\xi^\ast$ diverges. 
$(t^\prime,\mu) = (0,0)$,
with its special bipartite symmetry, is a unfractionalized 
``singular point'' in the space of spin-liquid insulators that we 
study. At this point, the wavefunction has a
power law decay
of spin correlations \cite{paramekanti}
$\sim (-1)^{x+y}/\vert \br\vert^\alpha$
with a non-trivial $\alpha \simeq 1.5$, and thus appears to 
be magnetically critical as well.

\section{Implications for the cuprates}
\label{cuprates}

Now the question arises --- what regime in parameter space
is relevant for the cuprates? One possibility is that there
are no fractionalized states in the vicinity of
the observed phases in real materials. An alternative
worth exploring in view of the success of
RVB wavefunctions in understanding the SC state \cite{paramekanti},
is that we take the insulating limit (hole doping $x\!\!\to\!\!0$) of
$\cP\vert BCS\ra$. 
Further, although $\cP\vert BCS\ra$ is not the ground state of the
for the half-filled Hubbard model at large-$U$, and
does not have long-range antiferromagnetic order, 
it is known to be an energetically competitive candidate state.
\cite{gros,paramekanti}

\begin{figure}
\begin{center}
\vskip-2mm
\hspace*{0mm}
\centerline{\fig{2.5in}{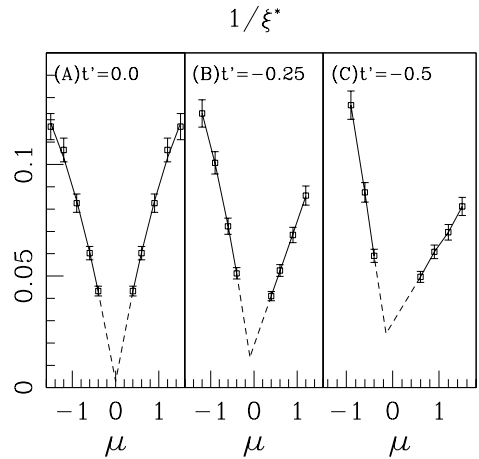}}
\vskip-6mm
\caption{Dependence of $1/\xi^\ast$ on
$\mu$ for (A) $t^\prime=0.0$, (B) $t^\prime=-0.25$, 
(C) $t^\prime=-0.5$. In panel (A), $1/\xi^\ast$ extrapolates to zero 
on approaching $(t^\prime,\mu)=(0,0)$ indicating a
unfractionalized ``singular point'' at the origin. 
For $t^\prime=-0.25,-0.5$ (B,C) the length scale $\xi^\ast$
appears finite everywhere.
}
\label{scaling}
\vskip-8mm
\end{center}
\end{figure}

Optimizing variational parameters 
$\mu,\Delta$ for the large-$U$ Hubbard model ($U\!\!=\!\!12, 
t^\prime\!\!=\!\!-1/4$),
we find $\Delta\!\!\simeq\!\!1.25$ while $-0.3\lesssim \mu \lesssim 0.3$ as 
shown in Fig.~\ref{phasediag}(B). For this region (the shaded ellipse in 
Fig.~\ref{phasediag}(A)) we conclude from Fig.~\ref{scaling}(B) that
$\xi^\ast \gtrsim 25$ lattice spacings. 

We now convert this to an estimate of the energy scale $E_v$
below which fractionalization is apparent.
We expect $E_v= \alpha J/(\xi^\ast)^z \le \alpha J/\xi^\ast$ 
which vanishes at the ``singular point'' with bipartite symmetry.
Here $J$ is the nearest-neighbor superexchange, $\alpha\equiv 
{\cal O}(1)$ is a dimensionless constant and the dynamical exponent 
$z \ge 1$.
For $\alpha\!\!=\!\!1$ and $J\!\!=\!\!1200 K$, the estimated 
$E_v \lesssim 50$K.  More concretely,
proximity to the unfractionalized 
`bipartite' point $(t'=0,\mu=0)$ can lead to a very 
small vison gap for fractionalized RVB states. Recently,
based on an ingenieous proposal of Senthil and Fisher \cite{senthil_2},
a flux trapping experiment \cite{moler}
was been carried out
to detect the vison in highly underdoped cuprates. 
However, these experiments did not see
the vison, which led to
an upper bound on the vison gap $\lesssim 150 K$. Our estimates
are consistent with this bound, and suggest
that $Z_2$ fractionalization, even if present in the cuprates, would
likely be apparent 
only at very low $T$ and cannot play a role in pseudogap anomalies. 

\section{application to numerical studies of Hubbard-like models}
\label{othermethods}

At this stage, we would like to emphasize the most important general 
outcome of our numerical results
which goes beyond the specific conclusions regarding the particular
wavefunction we have studied, and how it can be applied to numerical
studies of model systems \cite{lhuillier,imada,assaad}
to test for $Z_2$ fractionalization.
If a system is in a $Z_2$ fractionalized insulating phase, we have shown
that
the overlap of the wavefunctions with periodic and antiperiodic
boundary conditions imposed on the electrons must vanish 
exponentially as $\exp(-L/\xi)$ on an $L\times L$ system.
The length scale 
beyond which fractionalization is apparent, $\xi^\ast$, can be deduced 
from studying the finite size scaling of this overlap. These two
results are useful in detecting topological order
when the full low lying spectrum of a microscopic
Hamiltonian is not accessible so that the topological degeneracies are
not obvious even though one can work with
large system sizes. 

Consider, for example, Lanczos studies which
improve upon a definite trial wavefunction for any Hamiltonian
using a few Lanczos
iterations \cite{sorella}, or other Monte Carlo studies\cite{imada} 
which also use wavefunctions.
Let us imagine working on a cylinder (or torus), and 
starting with two different wavefunctions which differ 
in the boundary 
condition on the electrons. In a $Z_2$ spin liquid insulator, this
should provide us with the two topologically
degenerate ground states. However, if the
Hamiltonian has a conventional ground state however, we expect the
two resulting states to be identical.

Note that, if one has small system sizes, the two initial
states may have a large overlap even in a $Z_2$ spin liquid, 
and then would
lead to a single ground state. However, on a large enough system,
we expect the initial overlap would be exponentially small in a 
$Z_2$ spin liquid, so that one may recover the different topological sectors
in this manner.  Another check, apart from a vanishing overlap
of the different final states,
is that the energy difference between them
should vanish with increasing system size. If the spinon excitations
are gapped the energy difference is expected to decay exponentially 
in $L$, but
it would have a power law decay in the presence of gapless spinon
excitations.

\section{Concluding remarks}

In this paper, we have
analyzed the Gutzwiller projected d-wave BCS state at half-filling, 
and shown from analytic arguments and numerical results that
it generically describes a $Z_2$ fractionalized spin liquid except
at a special `bipartite' point where the vison is no longer 
well defined. We have pointed out the significance of this result
for the cuprate superconductors. As mentioned in the introduction,
there have been many recent examples of spin-liquid states reported
in numerical studies of $SU(2)$ spin models \cite{lhuillier}
and Hubbard-like models \cite{imada,assaad}. These models have been
identified as having spin liquid ground states based on the fact that 
the ground states do not appear to possess any simple broken symmetry 
patterns. The ideas and methods developed in this paper 
should be applied to these systems.
If they are shown to have topological order consistent
with that expected for a $Z_2$ spin liquid, they may provide us with
the first examples of $Z_2$ fractionalization in microscopic models
with full spin rotational symmetry.

\bigskip

\vspace{1cm}

\ni{\bf Acknowledgments:} 
We are grateful to T. Senthil for extensive discussions and
generously sharing his ideas and unpublished notes.
AP thanks A. Vishwanath for his
insights into $Z_2$ fractionalization during the course
of a related collaboration.
We also thank L. Balents, L. Capriotti, D.M. Ceperley, M.P.A. Fisher, 
E. Fradkin, A.J. Leggett and A. Melikidze for discussions. 
MR and NT gratefully acknowledge the 
hospitality of the Physics Department at University of Illinois and
support through DOE grant DEFG02-91ER45439 and DARPA grant N0014-01-1-1062.
AP was supported through NSF DMR-9985255 and PHY-07949 and grants from
the Sloan and Packard foundations. We acknowledge the use of computational 
facilities at TIFR including those provided by the DST Swarnajayanti 
Fellowship.


\begin{thebibliography}{999}
\bibitem{footnotelutt}
See Ref.\onlinecite{ashvin} for a discussion of crystal 
momentum quantum numbers of the different topological sectors 
in a $Z_2$ fractionalized phase, and how this is relevant for exact 
diagonalization studies which search for $Z_2$ spin liquids.
\bibitem{footnote2} 
We have checked that the particular projected wavefunction we use
has no long-range
magnetic or valence bond order and thus describes a 
translationally invariant spin-liquid.
\bibitem{footnote3}
We have checked that local spin 
correlations are identical in the thermodynamic limit for states with 
and without a vison. These states are thus locally identical 
and differ only in their global (topological) properties.
\bibitem{previous}
The same result was obtained by Wen \cite{wen} by analyzing stability of 
mean field solutions for spin models.
\bibitem{numerics} The Fourier transform of the d-wave
$\varphi(\bk)$ has singularities on the zone diagonal.
Thus, for technical reasons, on the square lattice we can only 
define $\vert V_x \ra$ and $\vert V_y \ra$. We have also studied
other geometries which permit all four states $(0,V_x,V_y,V_{xy})$; 
the results are consistent with those presented here.
\bibitem{footnotedelta}
This value of $\Delta$ was chosen as it is the value which
gives the minimum energy for the Hubbard model at half-filling. 
For more details on this, see Section \ref{cuprates} and 
Ref.\onlinecite{paramekanti}.
\bibitem{footnote4}
The factor of two is included in the asymptotic decay
definition since we are looking at overlap of states with two
visons in different directions.
\end{thebibliography}
\end{document}